\documentclass[aps,amsmath,amssymb,notitlepage,nofootinbib,twocolumn]{revtex4-1}

\usepackage[T1]{fontenc}
\usepackage[utf8]{inputenc}
\usepackage{xspace}

\usepackage{amsfonts}
\usepackage{bm}
\usepackage{physics}
\usepackage{slashed}

\usepackage{graphicx}
\usepackage{color,float}

\usepackage[pdftex,final]{hyperref}

\newcommand{\be}{\begin{equation}}
\newcommand{\ee}{\end{equation}}

\def\beq{\begin{equation}}
\def\eeq{\end{equation}}
\def\bea{\arraycolsep .1em \begin{eqnarray}}
\def\eea{\end{eqnarray}}
\def\s0#1#2{\mbox{\small{$ \frac{#1}{#2} $}}}
\def\0#1#2{\frac{#1}{#2}}

\renewcommand{\d}{\mathrm{d}}

\newcommand{\psib}{\bar{\psi}}
\newcommand{\hypergF}{\, _2F_1}

\newcommand{\intdx}{\int \d^d x \ }
\DeclareMathOperator{\sgn}{sgn}

\graphicspath{{./Figures/}}

\begin{document}

\author{Charlie Cresswell-Hogg}
\email{c.cresswell-hogg@sussex.ac.uk}
\author{Daniel F.~Litim}
\email{d.litim@sussex.ac.uk}
\affiliation{Department of Physics and Astronomy, University of Sussex, Brighton, BN1 9QH, U.K.}

\title{ Critical Fermions with Spontaneously Broken Scale Symmetry}


\begin{abstract}
We study 
relativistic   fermions in three euclidean dimensions with  four- and six-fermion interactions of the Gross-Neveu type.  In the limit of many fermion flavors, and besides the  isolated free fixed point, the theory   displays a  line of interacting ultraviolet fixed points. 
At the endpoint of the critical line, we establish that mass is generated  through the spontaneous breaking of quantum scale invariance.
Curiously, broken parity symmetry  is a prerequisite for the spontaneous generation of mass rather than a consequence thereof.
We also calculate   critical  exponents and find that hyperscaling relations are violated.
Further similarities with critical scalar theories, and 
implications for
conformal field theories and higher spin theories are discussed.
\end{abstract}

\maketitle

\section{Introduction}

Fixed points of the renormalisation group play a fundamental role in particle and statistical physics. At fixed points, theories  become scale- and possibly conformally invariant \cite{Polchinski:1987dy}, and correlation functions are  characterised by universal numbers \cite{cardy_1996}.
An intriguing scenario arises if quantum scale invariance is broken spontaneously. It leads to the appearance of a mass, which is not determined by the fundamental parameters of the theory. It has   been speculated that this type of mechanism may explain the Higgs particle as a light dilaton in   extensions of the Standard Model~\cite{Goldberger:2008zz,Bellazzini:2012vz,Csaki:2015hcd}. 

The spontaneous breaking of scale invariance has first been observed by Bardeen, Moshe, and Bander (BMB) in strongly coupled $3d$  $O(N)$ or $U(N)$ symmetric  scalar field theories~\cite{Bardeen:1983rv,Bardeen:1983st,David:1984we}. 
Besides the isolated Wilson-Fisher fixed point, 
the theory displays an asymptotically free 
line of ultraviolet (UV) fixed points at large $N$,  curtesy of an exactly marginal sextic scalar self-interaction~\cite{Bardeen:1983st,David:1984we,David:1985zz,Litim:2017cnl,Litim:2018pxe}.  At the tricritical endpoint, scale symmetry is broken spontaneously and hyperscaling relations are violated \cite{David:1985zz}, owing to a non-analyticity of the effective potential at vanishing field.
Subsequently, the phenomenon has  been observed  in multicritical bosonic theories~\cite{Eyal:1996da}, in finite $N$ extensions \cite{Litim:2017cnl}, and away from integer dimensionality \cite{Fleming:2020qqx}.  
 Further examples with spontaneously broken scale symmetry include 
  the Wess-Zumino model with $N=1$ supersymmetry~\cite{Bardeen:1984dx,Litim:2011bf,Heilmann:2012yf}, and models with  bosons or fermions coupled to a topological Chern-Simons term~\cite{Aharony:2012ns,Bardeen:2014paa,Moshe:2014bja,Sakhi:2019rfj}.

In this Letter, we study the spontaneous breaking of scale symmetry in a purely fermionic theory.
Our study is motivated by the recent discovery that Gross-Neveu-type $(\bar\psi\psi)^3_{\rm 3d}$ theories at large $N$ display a line of interacting UV fixed points and an isolated IR fixed point  \cite{Cresswell-Hogg:2022lgg},  very much like  bosonic $(\phi^2)^3_{\rm 3d}$  theories \cite{Bardeen:1983rv,Bardeen:1983st,David:1984we}, and with identical critical points and scaling dimensions.
It is conceivable that  this intriguing equivalence is rooted in a deeper connection between critical fermions and critical bosons. If so, we expect that the fermionic theory equally displays a version of spontaneous scale symmetry  breaking, in full analogy to the seminal  BMB phenomenon for  scalars. 
Here, with the help of renormalisation group and large-$N$ techniques, we demonstrate that this is  indeed the case.
Implications of our results for 3d fermion-boson equivalences, conformal field theory and higher spin theories are  indicated.

\section{Gross-Neveu Theory}

We briefly recall   fermionic quantum field theories in three euclidean dimensions with fundamental  four- and six-fermion interactions of the Gross-Neveu type~\cite{Gross:1974jv}. 
The corresponding classical action takes the form
\be\label{model}
S_{\rm f} = \! \!\int_x  
\Big\{ \psib_a( \slashed \partial +M)\psi_a + \frac G2 ( \psib_a \psi_a )^2 + \frac{H}{3!}  ( \psib_a \psi_a )^3 \Big\}.
\ee
where $\psi_a$ are four-component Dirac spinors, and summation over the index $a \in \{ 1, \dots, N \}$ is understood. 
The theory has a global $U(N)$ flavor symmetry. 
The theory (\ref{model}) is non-perturbatively renormalisable, and characterised by a line of ultraviolet (UV) fixed points in the limit of many fermion flavors $1/N\to 0$ \cite{Cresswell-Hogg:2022lgg}. In terms of the couplings $(m,g,h)$, which are the dimensionless counterparts of $(M,G,H)$ in the action, the line of fixed points is given by 
\begin{equation}\label{mghcrit}
m_*=0\,, \quad g_*=-\s012\,,\quad |h_*|\le h_*^{\rm crit}\,.
\end{equation}
At short distances, the mass term $\propto m$ and the four fermion (4F) interaction $\propto g$ are relevant operators, while the 6F coupling $h$ is exactly marginal.
Consequently,  small deviations $\delta m(\Lambda)$ and $\delta g(\Lambda)$ at the high scale $\Lambda$ and the value of the  6F coupling $h_*$ characterise UV-complete renormalisation group (RG) trajectories running from the UV to the IR.
 If the 6F coupling $h_*$ is taken to vanish, the theory is additionally invariant under a discrete parity symmetry \cite{ZinnJustin:1991yn,ZinnJustin:2002ru},
\be
\label{eq:discrete_symm}
\psi \to \gamma^5 \psi, \quad \psib \to - \psib \gamma^5\,.
\ee
It then follows that theories are either strictly massless $(\delta g>0)$, or  massive  $(\delta g<0$ or $\delta m\neq 0$) owing to the  dynamical or explicit breaking of parity symmetry. 

If the 6F coupling $h_*$ is non-zero, parity symmetry is absent and explicit mass terms or parity-odd (4n+2)F interactions are permitted.  
For $\delta m\neq 0$, mass is generated explicitly. For $\delta m=0$ and $\delta g<0$ with $|h_*|< h_*^{\rm crit}$, mass is generated dynamically through strong interactions, while for $\delta m=0$ and $\delta g>0$ with $|h_*|< h_*^{\rm crit}$, theories remain strictly massless, and parity symmetry ``emerges'' in the infrared \cite{Cresswell-Hogg:2022lgg}.
In this work, we investigate  the critical endpoint  $|h_*|= h_*^{\rm crit}$ to show that scale symmetry is broken spontaneously leading to a fermion mass without the breaking of any other symmetry. 

\section{Renormalisation group}
To uncover the phenomenon in question, we employ  functional renormalisation~\cite{Wetterich:1992yh,Ellwanger:1993mw,Morris:1993qb}, which has been used previously for  phenomena with spontaneous scale symmetry breaking  \cite{Litim:2011bf,Heilmann:2012yf,Litim:2017cnl,Litim:2018pxe}. Briefly, the method  proceeds by adding a Wilsonian cutoff term to the path integral representation of a quantum field theory, bilinear in the fields, which acts to integrate out successive momentum modes of the fields from the UV to the IR. By a Legendre transform, this defines a coarse-grained effective action $\Gamma_k$, dependent on the RG scale $k$, which interpolates between a classical action $S$ at some UV scale $k = \Lambda$ and a full quantum effective action $\Gamma$, the generating functional of one-particle-irreducible correlation functions, in the IR limit $k \to 0$.

The scale dependence of $\Gamma_k$, parametrised in terms of the dimensionless variable $t = \ln ( k / \Lambda )$, is governed by an exact functional identity
\be
\label{eq:wetterich}
\partial_t \Gamma_k = \tfrac{1}{2} {\rm STr} \left\{ \big[ \Gamma_k^{(2)} + R_k \big]^{-1} \cdot \partial_t R_k \right\},
\ee
known as the Wetterich equation, which is derived directly from the regulated partition function. The right hand side of this equation features a functional trace in position or momentum space, as well as a trace over all internal indices. The quantity $\Gamma_k^{(2)} + R_k$ stands for the exact inverse propagator of the regulated theory and includes the cutoff function $R_k$, which provides IR regularisation of the functional trace. UV regularisation is provided by the insertion of the scale derivative of the cutoff function. The cutoff term added to the action is bilinear in the fields, $\psib R_k \psi$, and in order that it respects the discrete symmetry~\eqref{eq:discrete_symm} we take the regulator proportional to $\slashed{q}$, in momentum space $R_k ( q ) = \slashed{q} \cdot r ( q^2 / k^2 )$, with a dimensionless function $r$ encoding its shape. The shape function should go to zero as $q^2/k^2$ grows large, allowing UV modes to propagate in the trace, and should become large as $q^2/k^2$ goes to zero, suppressing IR modes \cite{Litim:2000ci}. Additionally, its derivative should go to zero at large arguments. We consider the optimised cutoff~\cite{Litim:2001up,Litim:2002cf}
\be
\label{eq:r_opt}
r ( x ) = \left( \frac{1}{\sqrt{x}} - 1 \right) \cdot \Theta \left( 1 - x \right),
\ee
where $\Theta$ is the Heaviside step function. This choice allows for an analytical  evaluation of functional traces in~\eqref{eq:wetterich} and has been shown to improve the stability and convergence of approximations in a wide range of quantum and statistical field theories~\cite{Litim:2001fd,Litim:2003vp,Litim:2010tt}. Other choices constitute different RG schemes and we have checked that our central results are independent of the scheme.

In the large-$N$ limit, we solve the flow equation~\eqref{eq:wetterich} exactly using the ansatz
\be
\label{eq:gamma_ansatz}
\Gamma_k [ \psib, \psi ] = \intdx \Big\{ \psib_a \slashed{\partial} \psi_a + V_k \big( \psib_a \psi_a \big) \Big\}.
\ee
where we kept space-time dimension $d$ as a free parameter for now. 
It consists of a classical kinetic term and a scale dependent ``effective potential'' $V_k$, parametrising arbitrary interactions without derivatives built from the scalar combination $\psib_a \psi_a$. A virtue of the large-$N$ limit is that  tensor structures other than those already present in \eqref{eq:gamma_ansatz} are not generated along the RG flow~\cite{Gies:2010st,Jakovac:2013jua,Dabelow:2019sty}. Higher derivative operators will likewise remain absent from the action~\eqref{eq:gamma_ansatz} its form preserved under the evolution of~\eqref{eq:wetterich}, and the kinetic term remains unrenormalised and anomalous dimensions vanish~\cite{Jakovac:2013jua,DAttanasio:1997yph}. Hence, the closure of the ansatz~\eqref{eq:gamma_ansatz} under~\eqref{eq:wetterich} ensures that the theory can be solved exactly by solving the flow for $V_k$.

The flow equation for the function $V_k$ is obtained by inserting the ansatz~\eqref{eq:gamma_ansatz} into~\eqref{eq:wetterich} and projecting onto constant fields. It takes the form
\be
\label{eq:large-N_flow}
\partial_t v = - d \, v + ( d - 1 ) \, z \, v' - \frac{1}{1 + ( v' )^2},
\ee
which is written in terms of the dimensionless variables 
\be
z = k^{1 - d} \, \psib_a \psi_a,
\quad
v_k(z) = k^{-d} \, V_k ( \psib_a \psi_a )
\ee
with primes denoting partial differentiation with respect to the field variable $z$. Prior to taking the large-$N$ limit, each variable is additionally rescaled with a factor $4 N A_d$, where $A_d = S_{d-1} / [ d ( 2 \pi )^d ]$ with $S_n$ the surface of a unit $n$-sphere.
The first two terms on the right hand side of~\eqref{eq:large-N_flow} represent the classical scaling of $v$ and $z$, and are made explicit due to our choice of  variables, while the final term originates from the integrating out of quantum fluctuations, {\it i.e.} the right hand side of~\eqref{eq:wetterich}. Using the method of characteristics, the most general solution to \eqref{eq:large-N_flow} is found to be
\be
\label{eq:solutions}
z \cdot ( v' )^{1 - d} - F_d ( v' ) = G(v' e^t)\,,
\ee
where the function $G(x)=x^{1-d} z_\Lambda(x)-F_d (x)  $  is determined by  the boundary condition $v'_\Lambda(z)$ at $k=\Lambda$, and  with $z_\Lambda(v')$  the inverse function of $v'_\Lambda(z)$.
The function $F_d(x)$ can be expressed in terms of a Gaussian hypergeometric integral for arbitrary dimension, $F_d ( x ) = \frac{2}{d - 2} x^{2 - d} \, \hypergF \left( 2, 1 - \frac{d}{2}; 2 - \frac{d}{2}; - x^2 \right)$. For $d = 3$ it can be expressed in terms  of more elementary functions as
\be\label{eq:F3}
F_3 ( x ) =  -\frac{1}{x} \left(2+ \frac{x^2}{1 + x^2} + 3\, x \arctan x\right).
\ee 
Then, \eqref{eq:solutions} provides implicit solutions $z(v',t)$  which can be converted into explicit functions $v'(z,t)$ for any scale $k$ and all fields.

\begin{figure*}
\includegraphics[width=.75\linewidth]{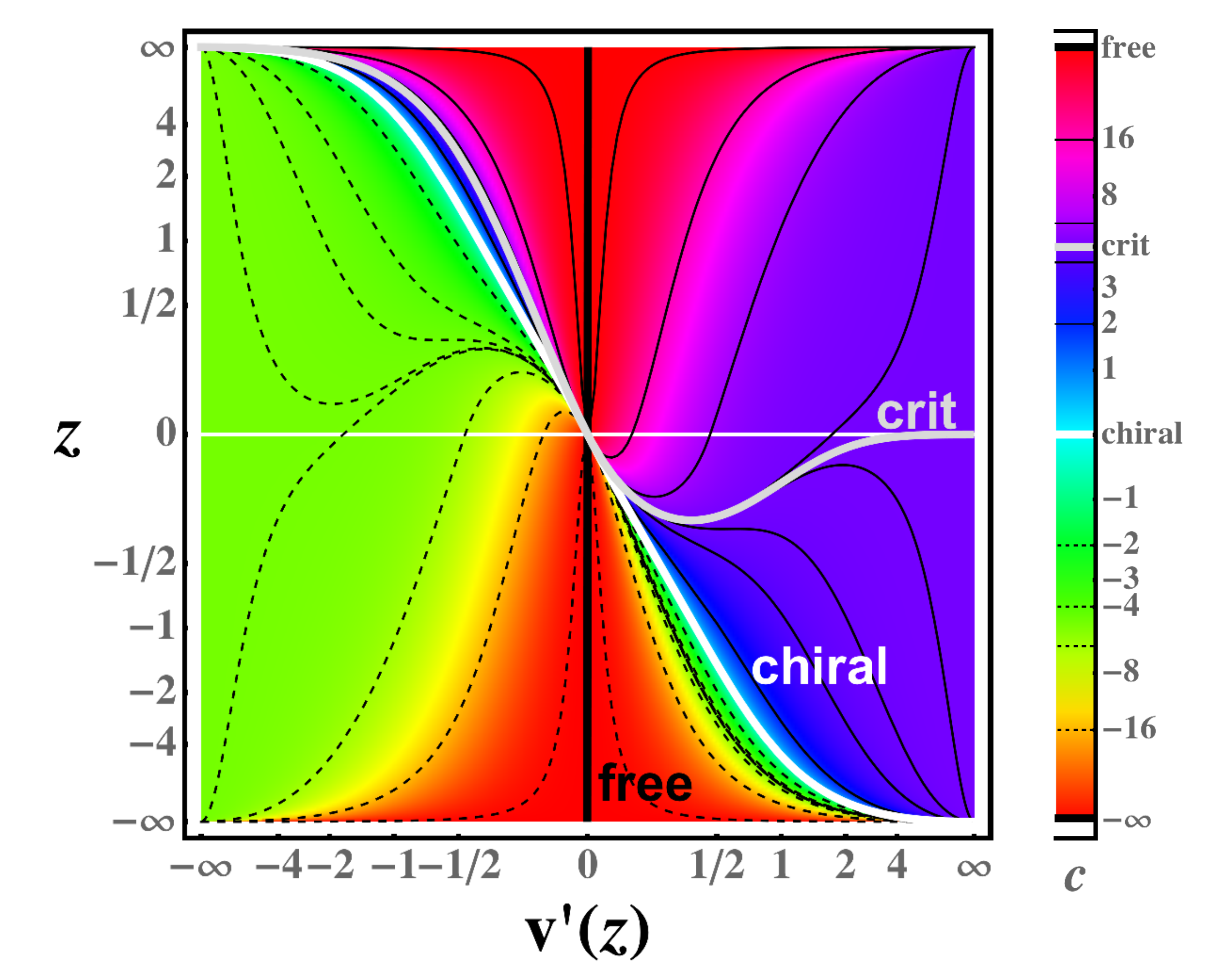}
\caption{Shown are all fixed point $v'(z)$ for all fields $z$ of the $(\bar\psi\psi)^3_{\rm 3d}$ theory  in the large-$N$ limit. Fixed points are characterised and colour-coded by the  parameter $c$. The chirally symmetric fixed point $(c_{\rm chiral}=0)$, the critical solution ($c_{\rm crit}=\frac32\pi$), and the free fixed point $(1/c_{\rm free}=0$)  are  highlighted by a  thick white, gray, and black line, respectively. A  further selection of solutions is shown by full ($c>0$) or dashed ($c<0$) black lines  to guide the eye. Global fixed points arise for $1/c=0$ (free) and for $|c|\in [0,c_{\rm crit}]$ (interacting). Axes are rescaled as $X \to X / ( 1 + | X | )$ for better display.}
\label{fig:FP_map}
\end{figure*}

\section{Critical points}
At a fixed point of the renormalisation group, the flow \eqref{eq:large-N_flow} vanishes identically, for all fields, and the theory becomes scale-invariant. 
To find all possible fixed points of the theory,
the right-hand side of \eqref{eq:solutions} must be $t$-independent, meaning that the  function $G$ can at best be a constant.
We therefore find a one-parameter family of integral curves $z(v' )$ with
\be
\label{eq:FP_solutions}
z \cdot ( v' )^{1 - d} - F_d ( v' ) = c, \quad c \in \mathbb{R},
\ee
which implicitly define all fixed points  \cite{Cresswell-Hogg:2022lgg}. 
A heat map of the solutions~\eqref{eq:FP_solutions} in three dimensions for all $c$ is shown in Fig.~\ref{fig:FP_map}.
Selected contours with constant $c$ are drawn to guide the eye.

Series expansions about vanishing field values make the nature of the fixed point solutions~\eqref{eq:FP_solutions} more explicit. For small $z$, we write
\be\label{eq:v_series}
v' ( z ) = m  +  g \, z + \frac{1}{2} \, h \, z^2 + \mathcal{O} ( z^3 ),
\ee
where $m$, $g$ and $h$ are the dimensionless mass, four-fermion and six-fermion couplings, respectively. On the other hand, from the fixed point solution \eqref{eq:FP_solutions} we find
\be\label{eq:seriesv'}
v'=-\frac{z}{2}+\frac{c}{8} z^{2}
-\0{4-c^2}{16} z^3
+\frac{5c(8-c^2)}{128}
z^4+{\cal O}(z^{5})\,.
\ee
From the solution  \eqref{eq:seriesv'} we confirm that the couplings in  \eqref{eq:v_series} are given by
the fixed point couplings \eqref{mghcrit}, with 
\be\label{eq:mgh}
h_* = \frac{c}{4}\,.
\ee
We observe that all fixed points have  vanishing mass parameter, identical nonzero 4F couplings, and a 6F coupling proportional to the free parameter $c$. Hence, the family of solutions described by~\eqref{eq:FP_solutions}  corresponds to a line of fixed points, continuously connected to the parity-even  Gross-Neveu theory by the 6F coupling $h$, which has become exactly marginal owing to  fluctuations. 

The line of fixed points does not extend indefinitely \cite{Cresswell-Hogg:2022lgg}, although the local field expansion~\eqref{eq:v_series} is blind to this fact. In fact, the range is limited to within 
\be\label{eq:range}
0\le |c|\le c_{\rm crit}\,,
\ee
where the critical value $c_{\rm crit}\equiv  \032 \pi$ relates to solutions~\eqref{eq:FP_solutions} which become singular at the origin of field space.
This result 
can be appreciated by considering the  large $|v'|$ limit. From \eqref{eq:F3} we have  $F_3( x)=- c_{\rm crit}-\s025x^{-5}+$ subleading ($x\gg 1$), which with \eqref{eq:FP_solutions} implies that almost all solutions have the asymptotic behaviour $v' \sim \sqrt{z}$. In other words,  the large $v'$ limit corresponds to the large field limit. For the critical value $c_{\rm crit}$, however,  the leading terms cancel out
and the subleading term dictates that
 large $v' \gg 1$  now relates to  small fields ($z\ll 1$) as
\be\label{eq:pole}
v'_{\rm crit}=(-\s052 z)^{-1/3}
+\text{subleading}\,,
\ee
instead of \eqref{eq:seriesv'}. The singular and non-analytic behaviour, which can be seen  in Fig.~\ref{fig:FP_map}  from the line marked ``crit'',  indicates  their  borderline nature beyond which the effective potential becomes either multivalued or ill-defined globally \cite{Cresswell-Hogg:2022lgg}. 
Incidentally, the small-field non-analyticity  is also similar to those  responsible for spontaneously broken symmetry      in critical scalar \cite{David:1985zz,Litim:2017cnl,Litim:2018pxe} or supersymmetric theories \cite{Litim:2011bf,Heilmann:2012yf}
 We conclude that the line of globally well-defined fixed points is indeed limited by  \eqref{eq:range}.

We close with a  comment on parity symmetry~\eqref{eq:discrete_symm}.
Invariance under parity is realised whenever $v'$ is an odd function of $z$. Since $F_3$ is itself an odd function, it follows  from~\eqref{eq:FP_solutions} that only the  $c = 0$  and the $1/c=0$ solution respects parity.
The  former corresponds to the well-known parity-even  UV fixed point without elementary six-fermion interactions~\cite{Gawedzki:1985ed,Gawedzki:1985jn,Rosenstein:1988pt,deCalan:1991km,Braun:2010tt,Braun:2011pp,Jakovac:2013jua,Jakovac:2014lqa,Ihrig:2018hho}. In Fig.~\ref{fig:FP_map},  the corresponding solution (white line)  is marked ``chiral''. 
The   limit $1/c\to 0$ relates to the free IR fixed point 
given by  the vertical line   (black)  marked ``free'' in Fig.~\ref{fig:FP_map}. 
For any other value of $c$, the fundamental theory at the UV fixed point is not parity invariant.

\section{Spontaneously Broken scale invariance}

Quantum field theories at free or interacting fixed points of the RG are scale invariant by definition. Ordinarily, this implies the absence of a mass scale. Following \cite{Heilmann:2012yf,Litim:2017cnl,Litim:2018pxe}, we explain how scale symmetry may nevertheless be broken spontaneously even though the theory is at an exact RG fixed point. To that end, we consider the quantity
\be\label{eq:Mass}
M_k= 
V_k'(\psib \psi )\Big|_{\psib \psi=0}\,,
\ee 
 which is extracted   from the function $V_k$ at vanishing field, also using \eqref{eq:solutions}. It relates to the physical fermion mass $M$   in the  limit $k\to 0$, irrespective of whether the theory is critical or not. At a fixed point, then, we have
\be\label{eq:mass}
M = \lim_{k \to 0} k \cdot m_*\,,
\ee
 where the dimensionless fixed point value $m_*$  is independent of $k$. Clearly, for finite $m_*$ masslessness follows trivially from \eqref{eq:mass}. Here, $m_*=0$ for any $|c|<c_{\rm crit}$.
However, this conclusion can be upset provided that $m_*$ diverges, which happens precisely at $|c|=c_{\rm crit}$.

To see the implications more explicitly, 
we express the family of fixed point solutions~\eqref{eq:FP_solutions} in terms of the dimensionful mass \eqref{eq:Mass}.
Taking the limit $k \to 0$ for theories at a fixed point, 
we  find a gap equation for the physical mass \eqref{eq:mass},
\be
\label{eq:BMB_phenom}
\left[ c - \frac{3 \pi}{2} \sgn ( M ) \right] M^2 = 0\,.
\ee
The gap equation depends on the parameter $c$ 
which is proportional to the critical six-fermion coupling. We emphasize that  the term $ \frac{3 \pi}{2} \sgn ( M ) $ in \eqref{eq:BMB_phenom}  originates from the non-analytical behaviour \eqref{eq:pole}.
For  any value of $c$ within the range \eqref{eq:range}, the prefactor is non-zero and the only solution to the gap equation~\eqref{eq:BMB_phenom} is that of a vanishing mass, 
\be M = 0\,,\ee 
in accord with \eqref{eq:seriesv'}  and the vanishing of \eqref{eq:mass}. Consequently, scale invariance at the fixed point remains intact.
However, at the boundary of \eqref{eq:range} where $c=\pm c_{\rm crit}$, the prefactor in \eqref{eq:BMB_phenom} vanishes identically, and the fermion mass is unconstrained and free to take any value,
\be M = \text{free parameter}\,,\ee
with $\sgn(M)=\sgn(c)$. 
 In consequence, scale symmetry is broken spontaneously, and precisely because of \eqref{eq:pole}. The result can also be interpreted as a version of dimensional transmutation \cite{Coleman:1973jx}, in that the role of a  dimensionless parameter, $c$, 
 is taken over by a dimensionful one, $M$, as a consequence of  quantum fluctuations. 
 However, we emphasize that the value of the spontaneously-generated fermion mass is not determined by the fundamental parameters of the theory and cannot be deduced from the non-analytic critical potential. These findings are in full qualitative agreement with the BMB phenomenon observed previously for critical sextic scalar theories \cite{Bardeen:1983st,David:1984we,David:1985zz,Litim:2017cnl,Litim:2018pxe}. In accordance with Goldstone's theorem, a single massless scalar mode should appear in the spectrum, the dilaton, related to the  generator of scale transformations.
We conclude that the critical endpoints points provide explicit examples of critical fermionic quantum field theories where mass is generated spontaneously.

\section{Violation of hyperscaling relations}

Critical exponents measure the response of macroscopic observables to changes in the microscopic parameters of a system close to criticality. 
Many fluids, magnets, or models in particle and condensed matter physics 
share the same behaviour at criticality described by universal numbers, such as the scaling exponent for the correlation length, $\nu$, or the anomalous dimension of the order parameter at criticality. Further critical exponents 
are linked to these two by scaling relations.
We begin by introducing the scaling exponent $\nu$ as
\be\label{eq:nu_def}
\xi \propto | r |^{-\nu}, \ r \to 0,
\ee
where $\xi$ is the correlation length and $r$ is a control parameter measuring the distance from the critical point. In the fermionic theory studied here, the correlation length is set by the physical fermion mass, $\xi \sim 1 / M$, which diverges at a critical point. 
We also introduce the specific heat exponent $\alpha$, defined by
\be\label{eq:alpha_def}
\chi \propto | r |^{-\alpha}, \ r \to 0.
\ee
In a thermal phase transition, $\chi$ relates to the specific heat, while its zero-temperature analogue, a ``control parameter susceptibility'', relates to the second derivative of the free energy with respect to $r$~\cite{Kirkpatrick:2015gia}.

The correlation length exponent  $\nu$, the specific heat exponent $\alpha$   and the space-time dimensionality $d$ are linked to each other by the hyperscaling relation
\be
\label{eq:hyperscaling}
d\, \nu = 2 - \alpha\,,
\ee
which generally holds true for all $d$ below the upper critical dimension~\cite{Kirkpatrick:2015gia}. Here, we demonstrate that the hyperscaling relation \eqref{eq:hyperscaling}  is violated.

Since the violation of~\eqref{eq:hyperscaling} originates from the  nonanalytic behaviour \eqref{eq:pole}, it  invalidates the extraction of critical exponents as  eigenvalues of RG beta functions. Therefore, we adopt  more elementary ideas \cite{David:1985zz} to identify the BMB scaling exponents at the critical endpoint. To that end, we solve the RG equation~\eqref{eq:large-N_flow} using the initial condition
\be
v ( z; t = 0 ) = m \, z + \frac12 \, g \, z^2 + \frac{1}{3!} \, h \, z^3
\ee
at $k = \Lambda$, see \eqref{eq:solutions}, and evolve the solution to the IR by taking $k \to 0$. We then obtain a gap relation at zero field, whose solution determines the physical mass $M$ for a given choice of the microscopic  parameters $m$, $g$, $h$. Fixing $g$ and $h$ to their critical values allows the scaling of $M$ with $m$ near a fixed point to be determined. In this manner we can extract the values of $\nu$ from~\eqref{eq:nu_def} as
\be
M \propto m^\nu, \ m \to 0^+\,.
\ee
in the different parameter regions.
In our setup, the gap relation takes the form of a transcendental equation $m = G ( g, h, \tilde M )$, where $G$ is a function of the initial parameters at the scale $\Lambda$ and the physical mass in units of this scale $\tilde M = M / \Lambda$. Without loss of generality, we take $h$ positive. Then an expansion in powers of $\tilde M$ yields
\bea
m &=& \left( 1 - \frac{g}{g_*} \right) \tilde M + \frac{1}{2 g_*^2} \left( h - \frac{g}{g_*} \, h_*^{\rm crit}  \right) \tilde M^2 \nonumber\\[1ex]
\label{eq:gap_expansion}
&&+ \left( \frac{h_*^{\rm crit}  h}{2 g_*^4} - 2 \frac{g}{g_*} \right) \tilde M^3 + \mathcal{O} ( \tilde M^4 ),
\eea
where $g_* = -\012$ denotes the value of the four-fermion vertex on the line of fixed points and $h_*^{\rm crit} = \038\pi$ is the six-fermion coupling at the critical endpoint.

We are interested in three regions of the parameter space spanned by $g$ and $h$, each giving rise to distinct critical behaviours. The first region is $g > g_*$, which leads to the free theory in the IR limit. In this regime~\eqref{eq:gap_expansion} dictates $m \propto M$ for $m \to 0$, reproducing classical scaling
\be\label{1}
m \propto M:\quad\quad\nu = 1
\ee
as it must. The second region are the points on the  critical line with  $g = g_*$ and $h < h_*^{\rm crit}$.
Upon setting $g$ to its critical value the linear term in~\eqref{eq:gap_expansion} drops out, leading to $m \propto M^2$ in the limit  $m \to 0^+$, with non-classical exponent 
\be\label{12}
m \propto M^2:\quad\quad\nu  = \frac12\,.
\ee
Lastly, we consider the critical endpoint point where  $g = g_*$ and $h = h_*^{\rm crit} $ to find the scaling relation. In this case both the linear and quadratic terms in~\eqref{eq:gap_expansion} vanish while the cubic term remains non-zero, resulting in $m \propto M^3$ and the critical exponent
\be\label{13}
m \propto M^3:\quad\quad \nu = \frac{1}{3}\,.
\ee
We emphasize  that a linearisation of the local RG beta functions would also have given the correct results at the free and the interacting fixed points \eqref{1} and \eqref{12}, respectively. However, at the critical endpoint,  the result from the local RG flows would have been \eqref{12} instead of \eqref{13}.
The failure of this standard method to capture the  scaling at the endpoints correctly is due to the fact that the local RG flow is unaware of the non-analyticity  \eqref{eq:pole}, which influences the scaling globally \cite{David:1985zz}. Finally, we note that it is not 
possible to make  the first three coefficients in the gap equation \eqref{eq:gap_expansion} vanish simultaneously, meaning that  the three  cases above cover all possibilities.

We now proceed to calculate the specific heat exponent $\alpha$.
The ordered phase of the system is characterised by a fermion condensate $Q = \langle \psib \psi \rangle$, which is proportional to the physical mass $M$ and is the conjugate variable to the bare mass $m$ in an expansion of the free energy. This latter point implies that the control susceptibility $\chi$ is proportional to  $\partial Q / \partial m$, up to an irrelevant prefactor. We then have $\alpha = 1 - \nu$,  provided  $\alpha > 0$. 
On the line of interacting fixed points, but away from the endpoints, we have $\nu = \012$, giving
\be
\alpha = \frac12\,.
\ee
This result is in accord with the hyperscaling relation~\eqref{eq:hyperscaling}. At the critical endpoints, however, we have $\nu = \013$ instead, leading to
\be\label{23}
\alpha = \frac{2}{3}\,.
\ee
We conclude that the hyperscaling relation~\eqref{eq:hyperscaling} is violated at the endpoints, where scale invariance is broken spontaneously. 

As a final remark, we note that our result is in quantitative agreement with the observed breaking of hyperscaling relations at the critical endpoint in scalar $(\phi^2)^3_{\rm 3d}$ models~\cite{David:1985zz}. As such, our work adds the new result that scaling exponents between the fermionic $(\bar\psi\psi)^3_{\rm 3d}$  and the bosonic $(\phi^2)^3_{\rm 3d}$ theories   agree at {\it all} UV or IR critical points, including at tricritical endpoints with spontaneously broken scale symmetry.

\section{Discussion and conclusions}

In this Letter, we have established for the first time that   purely fermionic quantum field theories may exhibit quantum critical points with spontaneously broken scale symmetry.
The fingerprint for scale symmetry breaking  are non-analyticities in the  effective  action  for the fermion bilinear $\psib \psi$ at a critical point (Fig.~\ref{fig:FP_map}).
The  spontaneously generated mass  breaks the conformal symmetry and becomes a new fundamentally free parameter of the theory.
Additionally, we have demonstrated that these non-perturbative effects
are equally responsible for the violation of hyperscaling relations. 

The underlying trigger for scale symmetry breaking in fermionic theories
is  similar to what has been observed  in critical scalar~\cite{Bardeen:1983rv,Bardeen:1983st,David:1984we,David:1985zz,Litim:2017cnl,Litim:2018pxe}   or  supersymmetric  models~\cite{Bardeen:1984dx,Litim:2011bf,Heilmann:2012yf}, even though  these theories  appear to be otherwise rather different. Their common denominator, however, 
is  the existence of a finite line of exactly marginal  deformations.
Whether the marginal direction arises out of an asymptotically free fixed point, 
such as in critical scalar or supersymmetric models \cite{Bardeen:1983rv,Bardeen:1983st,David:1984we,David:1985zz,Litim:2017cnl,Litim:2018pxe,Bardeen:1984dx,Litim:2011bf,Heilmann:2012yf}, or  an asymptotically safe one, such as in this work, is irrelevant from the viewpoint of scale symmetry breaking.

The marginality of $(\bar\psi\psi)^3$ interactions at large $N$
was previously noticed in~\cite{Gat:1991bf}, where it was also found that the 6F fixed point at the next-to-leading-order in $1/N$ is located in a region of instability.
This is consistent with  our work in that   the  Gat-Kovner-Rosenstein fixed point  \cite{Gat:1991bf}  relates  to one of the solutions with $|c|>c_{\rm crit}$, which are unphysical non-perturbatively (see Fig.~\ref{fig:FP_map}). This pattern  is    structurally mirrored   in critical scalar theories where a remnant of the  perturbative Pisarski fixed point~\cite{Pisarski:1982vz} is   similarily located  in the unstable region at the next-to-leading order, and thus unphysical non-perturbatively~\cite{Bardeen:1983rv,Bardeen:1983st}.

We also comment on  our results from the viewpoint  of fermion mass generation \cite{Cresswell-Hogg:2022lgg}.  A priori, fermion mass in the theory \eqref{model} can  be generated  both explicitly  and  dynamically,  with or without underlying parity symmetry.
What is new here is that  a fermion mass can  also arise spontaneously at an interacting  fixed point. 
Curiously,  for the theory \eqref{model}, it turns out that broken parity symmetry  is a prerequisite for the spontaneous generation of mass, and not   a consequence thereof.

At their critical points, our models become  conformal field theories (CFT). 
It would then be useful  to confirm  results using CFT techniques
and to understand whether  CFT three-point functions  
are modified at critical points where scale symmetry and hyperscaling relations are  broken 
\cite{Maldacena:2011jn,Maldacena:2012sf}. This is  also of interest for  higher spin gauge theories on AdS${}_4$, which relate through the AdS/CFT conjecture to critical bosonic \cite{Klebanov:2002ja} or critical fermionic theories \cite{Sezgin:2003pt} on the AdS boundary. 
While the duality is well-understood for parity-even boundary CFTs, it would be interesting to  understand whether higher spin duals can also be found for parity-odd boundary CFTs, with or without spontaneously broken scale symmetry.

Finally, the parallels between  fermionic $(\bar\psi\psi)^3_{\rm 3d}$ theories   
and bosonic $(\phi^2)^3_{\rm 3d}$  theories at large $N$ continue to be striking. 
Despite of their elementary differences on the level of the path integral, both  display equivalent  fixed points and scaling dimensions, and equivalent scenarios with spontaneously broken scale symmetry.
The evidence is   highly suggestive of a deeper link, perhaps similar to bosonisation dualities and holographic correspondences observed in Chern-Simons-matter theories~\cite{Maldacena:2011jn,Maldacena:2012sf,Aharony:2012ns,Bardeen:2014paa,Moshe:2014bja,Aharony:2018pjn}, or in the spirit of  large $N$ equivalences and orbifold reductions~\cite{Bond:2019npq}, which we take as natural directions for future work.

{\bf Acknowedgements.} This work is supported by the Science and Technology Facilities Council (STFC) by a studentship (CCH),  under the Consolidated Grant ST/T00102X/1 (DL), and was performed in part at the Aspen Center for Physics, which is supported by the National Science Foundation grant PHY-1607611 and by a grant from the Simons Foundation (DL).

\bibliography{fermionic_BMB_v10}
\bibliographystyle{mystyle2}

\end{document}